\documentclass[sigconf]{acmart}

\usepackage{booktabs} % For formal tables
\usepackage[utf8]{inputenc}

\usepackage{graphicx}
\usepackage{amsmath}
\usepackage[version=4]{mhchem}
\usepackage{siunitx}
\usepackage{longtable,tabularx}
\usepackage{amsfonts} 
\usepackage{url}
\usepackage{breakurl}
\usepackage{caption}
\usepackage{hyperref}

\usepackage{enumitem}
\setcopyright{acmcopyright}
\acmDOI{10.1145/3605098.3635976}

\acmISBN{979-8-4007-0243-3/24/04}

\acmConference[SAC'24]{ACM SAC Conference}{April 8 –April 12, 2024}{Avila, Spain}
\acmYear{2024}
\copyrightyear{2024}

\acmArticle{4}
\acmPrice{15.00}

\begin{document}
\title{Graph Learning-based Fleet Scheduling for Urban Air Mobility under Operational Constraints, Varying Demand \& Uncertainties
}
% \titlenote{Produces the permission block, and
% \subtitle{Extended Abstract}
% \subtitlenote{The full version of the author's guide is available asn
%   \texttt{acmart.pdf} document}
  
% \renewcommand{\shorttitle}{SIG Proceedings Paper in LaTeX Format}

\author{Steve Paul}
% \authornote{Dr.~Trovato insisted his name be first.}
\affiliation{%
  \institution{Department of Mechanical and Aerospace Engineering, University at Buffalo}
  % \streetaddress{P.O. Box 1212}
  \city{Buffalo} 
  \state{NY} 
  \country{USA}
  \postcode{14260}  
}
\email{stevepau@buffalo.edu}

\author{Jhoel Witter}
% \authornote{Dr.~Trovato insisted his name be first.}
\affiliation{%
  \institution{Department of Mechanical and Aerospace Engineering, University at Buffalo}
  % \streetaddress{P.O. Box 1212}
  \city{Buffalo} 
  \state{NY} 
  \country{USA}
  \postcode{14260}  
}
\email{jhoelwit@buffalo.edu}

\author{Souma Chowdhury}
% \authornote{Dr.~Trovato insisted his name be first.}
\affiliation{%
  \institution{Department of Mechanical and Aerospace Engineering, University at Buffalo}
  % \streetaddress{P.O. Box 1212}
  \city{Buffalo} 
  \state{NY} 
  \country{USA}
  \postcode{14260}  
}
\email{soumacho@buffalo.edu}
\thanks {\copyright\space2024 ACM. Personal use of this material is permitted. Permission from ACM must be obtained for all other uses, in any current or future media, including reprinting/republishing this material for advertising or promotional purposes, creating new collective works, for resale or redistribution to servers or lists, or reuse of any copyrighted component of this work in other works.}

\renewcommand{\shortauthors}{S. Paul et al.}

\begin{abstract}
This paper develops a graph reinforcement learning approach to online planning of the schedule and destinations of electric aircraft that comprise an urban air mobility (UAM) fleet operating across multiple vertiports. This fleet scheduling problem is formulated to consider time-varying demand, constraints related to vertiport capacity, aircraft capacity and airspace safety guidelines, uncertainties related to take-off delay, weather-induced route closures, and unanticipated aircraft downtime. Collectively, such a formulation presents greater complexity, and potentially increased realism, than in existing UAM fleet planning implementations. To address these complexities, a new policy architecture is constructed, primary components of which include: graph capsule conv-nets for encoding vertiport and aircraft-fleet states both abstracted as graphs; transformer layers encoding time series information on demand and passenger fare; and a Multi-head Attention-based decoder that uses the encoded information to compute the probability of selecting each available destination for an aircraft. Trained with Proximal Policy Optimization, this policy architecture shows significantly better performance in terms of daily averaged profits on unseen test scenarios involving 8 vertiports and 40 aircraft, when compared to a random baseline and genetic algorithm-derived optimal solutions, while being nearly 1000 times faster in execution than the latter. 
\end{abstract}




\maketitle

\section{Introduction}\label{sec:Introduction}
The concept of Urban Air Mobility (UAM) utilizes electric vertical take-off and landing (eVTOL) aircraft \cite{FAA_CON_OPS} to offer automated air transportation for passengers, cargo, and critical (e.g., air-ambulance) services, with a projected global market size of \$1.5 trillion by 2040 \cite{iii, jonas_jonas}. The economic viability of this new mode of transportation depends on the ability to operate a sufficiently large number of increasingly autonomous eVTOLs in any given market (i.e., achieve high penetration). This in turn demands safe airspace management and robust fleet planning solutions among others. 
More specifically, deploying a regional UAM network comprising scores of eVTOL aircraft requires an effective scheduling framework that can adapt to the unique demand patterns (that's different from general aviation) and aircraft and airspace constraints (distinct from other modes of regional/metropolitan transportation), while maximizing profitability and mitigating energy footprint \cite{Hwang2001}. These scheduling problems usually take the form of complex nonlinear Combinatorial Optimization (CO) problems, which can be addressed through classical optimization, heuristic search, and learning-based approaches. Approaches that provide local optimal solutions for small UAM fleet scheduling problems \cite{8569225, Shihab2020OptimalEF, 8901431} often present computational complexity that makes them impractical for online decision-making. By taking a multi-agent task planning perspective of the online fleet scheduling problem, in this paper, we propose a new reinforcement learning (RL) based solution. Moreover, this approach accounts for important problem complexities and constraints that are otherwise often overlooked by existing methods. These include constraints related to airspace corridors, aircraft charging, vertiport capacity, weather-induced uncertainties, and time-varying demand \cite{FAA_CON_OPS}. Air corridors in UAM are designated routes or paths in the airspace that are specifically allocated for the operation of the eVTOL.  Some of the associated technical challenges are summarized below. 

%The deployment of a successful UAM network requires a fleet scheduling framework that can address several challenges. These challenges include: 

\textbf{Dynamic environment:} The scheduling framework needs to consider real-time factors such as airspace and weather conditions, ground traffic, infrastructure availability, and demand uncertainty \cite{FAA_CON_OPS}. Thus, shorter time-scale, adaptive and robust planning is favored over fixed, deterministic, and/or day-ahead planning. It is also computationally challenging to resolve uncertainties in online planning.

\textbf{Conflict resolution:} 
The framework should facilitate sharing eVTOLs' state information and allow for trajectory and speed adjustments to ensure safe and optimal sharing of the airspace, which introduces additional constraints on decision-to-fly actions w.r.t. the route between any two vertiports \cite{FAA_CON_OPS}.

\textbf{Optimization \& resource allocation:} 
Optimizing (scarce) resource allocation, such as vertiport parking slots, charging stations, and air corridor capacity, is essential to prevent unnecessary delays in charging, takeoff, and other operations, which in turn impose additional constraints on scheduling actions \cite{FAA_CON_OPS}.

\textbf{Learning \& adaptation:} The capability of learning from past experiences, interactions, and feedback to adapt their decision-making strategies is needed. Learning algorithms facilitate agent adaptation to changing conditions, leading to improved efficiency and enhanced system performance by shifting the computational expense of training offline \cite{thipphavong2018urban}. Similar challenging characteristics can be found in other fleet planning and multi-robot transport or fulfillment planning problems.

Note that some of the above-stated challenges also appear in other critical fleet planning, multi-robot transport and fulfillment planning problems. To address these stated problem complexities and provide efficient online-executable policies for fleet scheduling, we explore the use of specialized Graph Neural Networks (GNNs) \cite{capam_mrta, paul2022scalable}. Our approach builds upon existing work in multi-robot task allocation, and through numerical experiments, demonstrates superior performance compared to standard RL-based and heuristic optimization-based solutions, as well as a feasible-random baseline.

\textbf{Related Work:}\label{sec:Related_works}
There is a small but growing body of work in UAM fleet planning, which has provided impetus for transitioning optimization and learning formalisms to advance this emerging concept. However, the majority of the existing work overlooks some of the important guidelines proposed by the US FAA \cite{FAA_CON_OPS} w.r.t. UAM airspace integration. For example, existing work usually lacks considerations for air corridors, range/battery capacity constraints, unforeseen events such as route closures due to bad weather or off-nominal events, or dysfunctional eVTOLs \cite{9482700, fernando2023graph, paul2022graph, 8901431}. Traditional methods such as Integer Linear Programming (ILP) and metaheuristics are not suitable for solving related NP-hard fleet scheduling problems in a time-efficient manner \cite{peng2021graph, miller1960integer, kamra2015mixed, Rizzoli2007, 7462285, zhang1999team, gen-2}. For perspective, here we consider hour-ahead planning, in order to enable enough capacity to adapt to varying demand and state of routes affected by weather and unanticipated aircraft downtime. Learning-based methods have in recent years shown promise for generating policies for CO problems with relatable characteristics \cite{Kool2019, barrett2019exploratory, Kaempfer2018LearningTM, 9750805, li2018combinatorial, nowak2017note, Tolstaya2020MultiRobotCA, Sykora2020, krisshnakumar2023fast, capam_mrta}. In their current form, however, they consider fewer complexities, tackle simpler problem scenarios or do not provide explicit capture of the problem-specific context. %Multi-agent Reinforcement Learning (MARL) approaches are another potential solution approach in this context, but are observed to have limited scalability and are affected by non-stationarity issues, making them less suitable for scheduling applications \cite{fernando2023graph, noureddine2017multi}. Moreover, for UAM fleet planning, a centralized decision-making is deemed more practical and (safety-wise) risk averse (as impact of imperfect communication are minimal at hour-ahead fleet planning scale), further limiting the potential suitability of decentralized MARL type approaches \cite{FAA_CON_OPS}.

We hypothesize that in order to address these complexities in UAM fleet scheduling or related problems, in a manner that would be both generalizable across scenarios/environments, a suitable representation of the problem space is needed. Subsequently, we need to identify a neural architecture that can efficiently operate on this representation to provide reliable solutions with a small optimality gap. To this end, firstly we explore the use of a graph abstraction of the eVTOL state and vertiport state space, which is amenable to adding or changing problem/environment features. Secondly, we create a lightweight simulation environment that incorporates the modeling of the principal constraints and uncertainties. Finally, we propose a new graph neural net (GNN) type policy architecture to operate on the graph abstraction of the problem and utilize the simulation environment to generate hour-ahead sequential actions for eVTOLs over a generic (12-hr) day of operation, acting in the role of a centralized planner. The policy network combines a Graph Capsule Convolutional Neural Network (GCAPCN) \cite{Verma2018} to encode vertiport and eVTOL state information, a Transformer encoder network to incorporate time-series data on demand and fare (similar to \cite{chuwang2022forecasting}), and a feedforward network to encode passenger transportation cost. Additionally, a Multi-head Attention mechanism is used to fuse the encoded information and problem-specific context, to compute the sequential actions \cite{Verma2018, VaswaniSPUJGKP17, Kool2019}.

\textbf{The Main Contributions} 
of this paper can thus be summarized as \textbf{1)} Formulating the UAM fleet scheduling problem as a Markov Decision Process (MDP), and architecting a centralized encoder-decoder policy network, where the state of the UAM network (vertiports and aircrafts) is embedded by a special Graph Neural Network (GNN), with the demand, passenger fare, operating cost and air corridor availability information processed by different Context encoders and then concatenated. The scheduling actions are computed using a Multi-head Attention (MHA) based action decoder that is fed by the GNN and context. \textbf{2)} Integrating the encoder-decoder policy network with a new simulation environment that models the daily operation of 40 eVTOLs across 8 vertiports, associated uncertainties, and airspace/aircraft constraints, and provides data on demand, fare, and operational costs. \textbf{3)} Training the encoder-decoder network via policy gradient techniques and demonstrating its ability to generalize across unseen scenarios and uncertainties. \textit{We expect that, with problem-specific design of the context portion, this policy architecture can generalize to a wider range of multi-agent/vehicle/robot scheduling problems with similar complex characteristics, namely graph-abstractable task/resource space, time-varying and uncertain environment properties, and the need for sequential actions that satisfy a large set of physical and operational constraints.} %\textcolor{red}{This can be achieved by the appropriate graph formulation of the state information and the corresponding change in the context portion}.}  %\textbf{4)} Developing a framework with a simulated environment for UAM fleet scheduling operations, incorporating time-varying demand, passenger cost, uncertainties, and constraints guided by FAA guidelines \cite{FAA_CON_OPS, Vertiport_design_FAA}.

\textbf{Paper Outline:} The next section describes the UAM fleet scheduling problem and its MDP formulation. Section \ref{sec:Proposed_solution} explains the proposed solution, including the state encoding and action decoding. Section \ref{sec:experimental_evaluation} covers training and experimental evaluations, and Section \ref{sec:conclusion} presents concluding remarks.

 \begin{figure}[!ht]
    \centering
    \includegraphics[width=1.0\columnwidth]{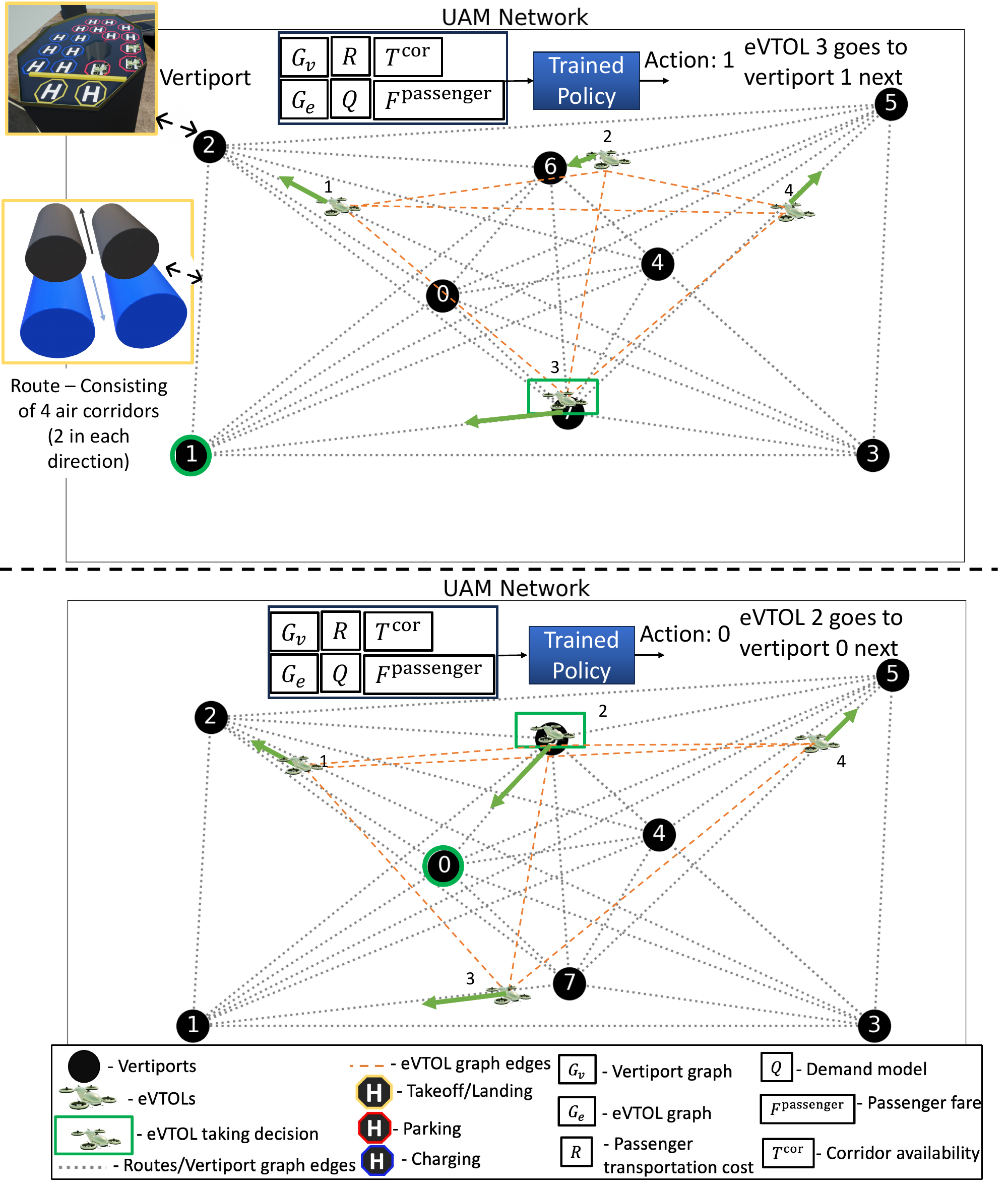}
    \caption{The trained policy implemented on the UAM network: The two images show two consecutive decision-making instances. For clarity of illustration, only 4 eVTOls are shown.}
    \label{fig:decision-making} 
\end{figure}

\vspace{-.1cm}
\section{Problem Description \& Formulation}\label{sec:Problem_description_formulation}
%%%%%%%%%%%%%%%%%%%%%%%%%%%%%%%%%%%%%%%%%%%%%%%%%%%%%
% \begin{itemize}
%     \item X hour ahead scheduling (this will be from learned model)
%     \item After a cancellation of flight, reschedulling is required - this should be mentioned in the testing
%     \item Mention that the features, complexities, and constraits considered are based on the FAA conops document, and the one from Boeing
% \end{itemize}

% \textcolor{red}{Write about rebalancing (space-driven or demand-driven)}

A UAM network includes vertiports and eVTOLs (or aircraft) for passenger transportation, with vertiports providing take-off/landing spots, charging spots, parking areas, and terminals for boarding and departing. 
We consider a concept of operations (ConOps), where every vertiport is connected to each other (i.e., a \textit{fully} connected vertiport network \cite{9925782}) by a route that comprises air corridors, as shown in Fig. \ref{fig:decision-making}. For safe operation, eVTOLS can only use these air corridors for flight. Since UAM is still an emerging concept, the exact connectivity structure of a UAM network is not yet established. Hence we assumed a fully connected network, i.e., where all the nodes are connected to each other. %, while in a connected network, there exists a direct or indirect (multi-hop) connection between two nodes. 
The corridors have regulations regarding the required distance gap for safe navigation. Here, we consider the corridors to be straight tube-shaped air columns. We consider there to be 4 air corridors between two vertiports with two corridors for each direction. The UAM network is defined as involving $N$ vertiports, and $N_{K}$ number of eVTOLs, with each eVTOL having a maximum passenger seating capacity of $C$ (=4) and maximum battery capacity of $B_{\text{max}}$ (=110kWh). Let $V$ and $K$ be the set of all vertiports and eVTOLs, respectively. Each vertiport $i \in V$ has a maximum number ($C^{\text{park}}_{\text{max}}$=10) of eVTOLs it can accommodate at a time and the total number of charging stations it contains ($C^{\text{charge}}_{\text{max}}$=6).  Some of the vertiports that do not have a charging facility, called vertistops ($V_{s} \subset V$), are only used for landing/take-off, and passenger boarding and have parking spaces. For computing the cost of transportation, we define $R_{i,j}$ to be the cost of transporting a passenger from vertiport $i$ to $j$. The travel demand between vertiports is modeled based on real data as explained in Section \ref{subsec:Demand_model}. Each vertiport $i \in V$, has an expected take-off delay $T^{\text{TOD}}_{i}$ which will affect every take-off from the vertiport during an episode. Here, an episode refers to the UAM operation for a specific time period.

We make the following assumptions for setting up this problem:
    \textbf{1)} A single service provider runs the entire UAM network, and full observation of the states of aircraft and vertiports in the network is available to the central agent making the scheduling decisions, which is reasonable under current communication capabilities in urban areas, and given that planning occurs at 15-60 min time scales. 
    \textbf{2)} Every eVTOL can commute between any two vertiports if it has enough battery charge (considering a safety margin) for the commute.
    \textbf{3)} The resistive loss of the batteries is negligible.
    \textbf{4)} The state information of eVTOLs and vertiports is always accessible for decision-making, which is crucial for ensuring passenger safety in aviation applications.
    \textbf{5)} An estimate of the probability of unplanned technical grounding of eVTOLs is available.
    \textbf{6)} While route closure on a given day of operation is not known apriori, an estimate of the probability of route closure (due to factors such as bad weather) is however available.
    \textbf{7)} The expected take-off delay ($T^{\text{TOD}}_{i}, i \in V$) at every vertiport can be estimated and is considered to be less than 6 minutes.
 We consider a probability of route closure between two vertiports as $P^{\text{closure}}_{i,j} (\leq 0.05), \forall~ i,j \in V$. Once the route is closed, there will not be any more commutes between these two vertiports during the rest of the episode. We consider the probability of an eVTOL to become dysfunctional as $P_{k}^{\text{fail}} (\leq .005), \forall~ k \in K$ during an episode. In every episode, we consider a randomly assigned take-off delay (of $<30$ mins) for any vertiport-$i$, based on a Gaussian distribution with mean $T^{\text{TOD}}_{i}$ and a standard deviation of 6 minutes. We assume this probability distribution of take-off delay to be known prior to the scheduling. While these uncertainties are modeled for realism, and seeded with prescribed values due to lack of historical data in this regard, these values can be readily tweaked once data (or forecasts) become available. Passenger pricing for a journey is based on operational cost per passenger and the demand for the trip. Upon reaching a vertiport, each eVTOL is fully charged before proceeding to the next destination (excluding vertistops) or before being parked in an available spot if it is to be idle.

The planning objective is to maximize the daily profit by optimizing the schedule of eVTOL flights between vertiports to meet travel demand. Each decision-making instance involves assigning an eVTOL (currently at a vertiport) to another vertiport to fly to, or instructing it to remain idle for another 15 mins. Factors such as demand, battery charge, and operational costs are taken into account. The current implementation focuses on 4-hour-long episodes (for computational ease of training) between 6 am to 6 pm without an end-of-episode constraint. Each episode is independent, starting with a random number of fully charged eVTOLs at each vertiport, subject to vertiport capacity constraints. The passenger demand model, eVTOL battery model, and optimization formulation of the problem are further discussed below.
\vspace{-.1cm}

\subsection{Demand Model}\label{subsec:Demand_model}
 Passenger demand modeling is used to generate stochastic passenger requests between different vertiports. The demand is based on forecasted data from \cite{Administration}, and we assume to know the expected demand for each hour during daily operating hours. The demand between two vertiports $i$ and $j$ at a time $t$, $Q(i,j,t)$, mimics the demand patterns of a major city's subway system, where certain stations close to commercial hubs and workplaces experience higher traffic. In our scenario, a subset of vertiports, $V_{B} \subset V$, is designated as high-demand vertiports. The demand between vertiports in $V_{B}$ is higher compared to those in $V - V_{B}$. We consider two peak hours: 8.00-9.00 am ($T^{\text{peak}1}$) and 4.00-5.00 pm ($T^{\text{peak}2}$). Vertiports in $V_{B}$ experience peak demand during both hours, resembling the morning and evening rush hours for commuting between home and workplace. This represents high demands from vertiports in $V - V_{B}$ to those in  $V_{B}$ during $T^{\text{peak}1}$, and vice versa in $T^{\text{peak}_2}$.  
 % \begin{equation}
 %     Q(i,j,t) = 
 %     \begin{cases}
 %    \mathcal{N}(100, 10) \ i \in V_B, j \in V_B i\neq j, \ t =T^{\text{peak}_1} \ \\ or \ t=T^{\text{peak}_2} \\
 %    \mathcal{N}(100, 10) \ i \in V-V_B, j \in V_B, i\neq j, \ \\ t =T^{\text{peak}_1} \ \\
 %    \mathcal{N}(100, 10) \ i \in V_B, j \in V-V_B, i\neq j,  \ \\ t=T^{\text{peak}_2} \\
 %    \mathcal{N}(50, 10) \ i \in V_B, j \in V_B, i\neq j, \ t \neq T^{\text{peak}_1} \ \\ and \ t \neq T^{\text{peak}_2} \\
 %    \mathcal{N}(30, 5) \ \text{otherwise}
 %     \end{cases}
 % \end{equation}
 % where $\mathcal{N}(\mu, \sigma)$, represents a normal distribution with mean $\mu$ and standard deviation $\sigma$.
 \vspace{-.1cm}
\subsection{eVTOL model}\label{subsec:eVTOL_model}
The eVTOL vehicle model considered here is the same as in \cite{paul2022graph} (City Airbus eVTOL aircraft), having a maximum cruise speed of $74.5 \ \text{mph}$, maximum passenger capacity of $4$, and a maximum range of $50$ miles The operating cost of the vehicle is about $\$0.64$ per mile \cite{Shihab2020OptimalEF}. We also assume that for every eVTOL $k$, a downtime probability of being dysfunctional is given by $P_{k}^{\text{fail}}$; this is based on work on aircraft predictive maintenance \cite{schoppmann2022operation}. Once an eVTOL become dysfunctional, it can no longer be in service for the remainder of the episode. The battery model is considered to be the same as that in \cite{paul2022graph}, which consists of a maximum capacity of $B_{\text{max}}$= 110 kWh. If $B^{k}_{t}$ is the battery charge of eVTOL $k$ at time $t$, and assuming the eVTOL travels from vertiport $i$ to $j$, the charge for the next time step $B^{k}_{t+1}$ can be computed as $B^{k}_{t+1} = B^{k}_{t} - B^{\text{charge}}_{i,j}$ Here $B^{\text{charge}}_{i,j}$ is the charge required to traverse between vertiports $i$ and $j$. The batteries are charged at vertiports with a charging rate of 150 kW.
\vspace{-.1cm}
\subsection{Passenger Fare 
 \& Electricity Pricing Models}
The passenger price consists of a fixed base fare, $F^{\text{base}}$, of $\$5$, and a variable fare, $F^{\text{passenger}}$. The variable fare between two vertiports for a passenger at a time $t$ is computed as a function of the demand profile, $Q(i,j,t)$, and the operational cost, $R_{i,j}$ (between $i$ and $j$), expressed as $F^{\text{passenger}}_{i,j,t} = R_{i,j} \times Q_{\text{factor}}(i,j,t)$. Here $Q_{\text{factor}}(i,j,t) = \text{max}(\text{log}(Q(i,j,t)/10), 1)$.
This is a hand-crafted function that accounts for demand in passenger pricing. A constant electricity pricing, $\text{Price}^{\text{elec}}$ of \$0.2/kWh is used here \cite{electric_price}.
\vspace{-.1cm}
\subsection{Optimization Formulation}\label{subsec:optimization_formulation}

The objective function to be maximized in the fleet scheduling problem is the daily profit, given by the difference of the earned revenue, and the operating and electric-charging costs. Uncertainties due to route closures, eVTOL malfunctions, and expected delays affect this objective function. Here, the decision variables are the destination vertiports ($V^{\text{end}}_{k,l}$ in the following paragraph) of all the eVTOLs for all journeys.  Hence, we are presented with a stochastic Integer Nonlinear Programming (INLP) problem. 

Consider a time period $T$ (with $|T|$ hours), with a start time of $T^{\text{start}}$ and end time of $T^{\text{end}}$. We consider cases where $T^{\text{end}}$ $<$ 6.00 pm, and $T^{\text{start}}$ = $T^{\text{end}} - |T|$, and $T^{\text{start}}$ $\geq$ 6.00 am. For every eVTOL, $k \in V_{e}$, let $S^{\text{jour}}_{k}$ be the set of journeys taken during time period of $T$; this could be of different length for different eVTOLs. Let $N_{i,l}^{\text{passengers}}$ be number of passengers transported during the trip, $l \in S^{\text{jour}}_{k}$ by eVTOL $k$. Let $B^{k}_{l}$ be the battery charge of eVTOL $k$ just before its $l$'th journey, $V^{\text{start}}_{k,l}$ and $V^{\text{end}}_{k,l}$ be the respective start and end vertiports of eVTOL $k$ during its $l$'th journey, $T^{\text{takeoff}}_{k,l}$ and $T^{\text{landing}}_{k,l}$ be the corresponding takeoff time and landing time.

Therefore, the total cost of transportation, ($\text{Cost}^{\text{Oper}}$) and charging, ($\text{Cost}^{\text{Elec}}$), and the revenue during $T$ are given by:
\begin{equation}\label{eq:cost_oper}
% \footnotesize
    \text{Cost}^{\text{Oper}} =  \sum_{k \in K} \sum_{l \in S^{\text{jour}}_{k}} N_{i,l}^{\text{passengers}} \times R_{i,j} , \ i = V^{\text{start}}_{k,l}, \ j = V^{\text{end}}_{k,l}
\end{equation} \vspace{-0.8cm}

\begin{equation}\label{eq:cost_elec}
% \footnotesize
\text{Cost}^{\text{Elec}} = \sum_{k \in K} \sum_{l \in S^{\text{jour}}_{k}} \text{Price}^{\text{elec}} \times B^{\text{charge}}_{i,j}, \ i = V^{\text{start}}_{k,l}, \ j = V^{\text{end}}_{k,l} 
    % C^{C} = \sum_{jk \in J_{k}}  \sum_{k \in K}  \text{Price}^{elec}}_{t} g_{i,j,k}, \ i = Loc^{k}_{\text{start}}(jk), \ j = Loc^{k}_{\text{end}}(jk)
\end{equation} \vspace{-0.4cm}
\begin{align}\label{eq:revenue}
% \footnotesize
\text{Revenue} = \sum_{k \in K} \sum_{l \in S^{\text{jour}}_{k}} N^{\text{passengers}}_{i,l} \times F^{\text{passenger}}_{i,j,t}, \\ \nonumber \ i = V^{\text{start}}_{k,l}, \ j = V^{\text{end}}_{k,l}, \ t =T^{\text{takeoff}}_{k,l} 
    % R^{T} = \sum_{jk \in J_{k}} \sum_{k \in K} N^{P}_{k}(jk) P_{ijt}, \ i = Loc^{k}_{\text{start}}(jk), \ j = Loc^{k}_{\text{end}}(jk)
\end{align}

Therefore, the objective function can be expressed as:
\begin{equation} \label{eq:objective_function}
% \footnotesize
    \max z = \text{Revenue} - \text{Cost}^{\text{Oper}} - \text{Cost}^{\text{Elec}} 
\end{equation}
We must also satisfy the following operational constraints:
\begin{align}
% \scriptsize
    N_{i,l}^{\text{passengers}} = \min(C, Q_{\text{act}}(i,j,t)), 
    i = V^{\text{start}}_{k,l}, \ j = V^{\text{end}}_{k,l}, 
  \nonumber \\  \ t =T^{\text{takeoff}}_{k,l} \ \forall~ l \in S^{\text{jour}}_{k} \label{c1} \\
    B^{k}_{T^{\text{takeoff}}_{k,l}} > B^{\text{charge}}_{i,j},\ i = V^{\text{start}}_{k,l}, \  j = V^{\text{end}}_{k,l}, \ \forall~ k \in K, \ \forall~ l \in S^{\text{jour}}_{k} \label{c2} \\
    C^{\text{park}}_{i,t} \le C^{\text{park}}_{\text{max}}, \ \forall~ i \in V_{v}, \ \forall~ t \in \left[T^{\text{start}}, T^{\text{end}}\right]  \label{c3} \\
    V^{\text{end}}_{l,k} \neq i, \ {\text{if}} \ A^{i, V^{\text{end}}_{l,k}} =0 \ \forall~ i \in V_{v}, \ \forall~ k \in K, \ \forall~ l \in S^{\text{jour}}_{k}\label{c4}
\end{align}
Here $B^{\text{charge}}_{i,j}$ is the amount of charge required for an eVTOL to fly from vertiport $i$ to $j$, $Q_{\text{act}}(i,j,t)$ is the actual demand sampled from $Q$, $P^{\text{closure}}$ represents the probability matrix for route closure, and $A$ is a binary matrix such that $A_{i,j} = A_{j,i}=0$, if the route between vertiports $i$ and $j$ is closed. Equations \ref{eq:cost_oper}, \ref{eq:cost_elec}, and \ref{eq:revenue} are used to compute the objective function (Eq. \ref{eq:objective_function}). Equation \ref{c1} is used to compute the number of passengers transported by eVTOL $k$ on its $l$'th journey. Constraint in Eq. \ref{c2} ensures that eVTOL $k$ has enough charge before taking off for its $l$'th journey. Constraint in Eq. \ref{c3} ensures that the number of eVTOLs parked in every vertiport does not exceed its maximum capacity. Constraint in Eq. \ref{c4} ensures that infeasible vertiports are not chosen during a decision-making instance. Later on, a Genetic Algorithm is applied on this exact optimization formulation to compute optimal solutions for comparisons with the learning based solutions.
\vspace{-.1cm}
\subsection{MDP Formulation}\label{subsec:MDP+formulation}
% \begin{itemize}
%     \item Why model as MDP and not Markov games?
%     \item Why centralized decision-making?
% \end{itemize}

In this work, we formulate the fleet scheduling problem as an MDP, where actions are computed sequentially for each eVTOL during its decision-making instance ($t \in [T^{\text{start}}, T^{\text{end}}]$). At each time step, an action is assigned to each eVTOL based on the current state of the vertiport network, which contains all necessary information for decision-making. Unlike other multi-agent approaches in smaller UAM settings, here we impose high safety standards \cite{FAA_CON_OPS}. Therefore, full state information is essential for decision-making, leading us to adopt a centralized decision-making scheme.

\textbf{Graph Formulation for UAM vertiport Network:}
The UAM vertiport network is expressed as a graph, $G_{v} = (V_{v},E_{v},A_{v})$, where $V_{v}$ (=$V$) represents the set of nodes or vertiports; in this case, $E_{v}$ represents the set of edges between the nodes, and $A_{v}$ represents the adjacency matrix of the nodes. Since we consider a route closure probability $P^{\text{closure}}$, we compute the weighted adjacency matrix as $A_{v} = (\mathbf{1}_{N \times N} - P^{\text{closure}}) \times A$, where $A$ is a matrix representing the route closure such that if $A_{i,j}$ is the route between nodes $i$ and $j$ ($i,j \in V_{v}$), then $A_{i,j} = A^{j,i} = 0$, if the route is closed; else it is equal to $1$. Here, the properties, $\delta^{t}_{i}$, of each node $i \in V_{v}$ at the time step $t$ are: \textbf{1)} the x-y coordinates of the node/vertiport ($x_{i}, y_{i}$),   \textbf{2)} the number of eVTOLs that are parked at vertiport $i$ at time $t$, $C^{\text{park}}_{i,t}$, \textbf{3)} the earliest time at which a charging station is available $T^{\text{charge}}_{i}$, \textbf{4)} the expected take-off delay $T^{\text{TOD}}_{i}$, and \textbf{5)} a binary number $I^{\text{vstop}}_{i}$ which takes a value of 1 if the node is a vertistop. Hence, $\delta^{t}_{i} = \left[x_{i}, y_{i}, C^{\text{park}}_{i,t},T^{\text{charge}}_{i}, T^{\text{TOD}}_{i}, I^{\text{vstop}}_{i}\right]$,  $\delta^{t}_{i}$ $\in$ $\mathbb{R}^{6}$. 

\textbf{Graph Formulation for eVTOLs Network:} The state of eVTOLs is also represented as a graph, $G_{e} = (V_{e},E_{e},A_{e})$, where $V_{e}$ represents the set of eVTOLs, $E_{e}$ represents the set of edges between the nodes, and $A_{e}$ represents the adjacency matrix. We consider $G_{e}$ to be fully connected. Each node $k$ $\in$ $V_{e}$ has its time-varying node properties, $\psi^{t}_{k}$. The properties of a eVTOL node are \textbf{1)} the coordinates of the destination vertiport ($x^{d}_{k}, y^{d}_{k}$), \textbf{2)} the current battery level as a fraction $B_{t}^{k}$, \textbf{3)} the next flight time $T^{\text{flight}}_{k}$, \textbf{4)} the next decision-making time $T^{\text{dec}}_{k}$, and \textbf{5)} the probability of failure $P_{k}^{\text{fail}}$. Therefore, $\psi^{t}_{k} = \left[x^{d}_{k}, y^{d}_{k}, B_{t}^{k}, T^{\text{flight}}_{k}, T^{\text{dec}}_{k}, P_{k}^{\text{fail}}\right]$, $\psi^{t}_{i}$ $\in$ $\mathbb{R}^{6}$.

\textbf{State Space:}
The state information that will be used for computing the action at a time step $t$ consists of:
\textbf{1)} Vertiport graph $G_{v}$, \textbf{2)} eVTOLs graph $G_{e}$, \textbf{3)}  Demand profile $Q$, \textbf{4)} Passenger fare $F^{\text{passenger}}$, \textbf{5)} Cost of per passenger transportation $R$, and \textbf{6)} Time at which it's safe to launch an eVTOL to the corridors $T^{\text{cor}} \in \mathbb{R}^{N \times N \times 2}$. It should be noted some of the state variable updates such as $T^{\text{flight}}, T^{\text{dec}},T^{\text{charge}}_{i}, T^{\text{TOD}}_{i}$, etc. are not explicitly presented in this paper due to space constraints; however, they have been appropriately implemented programmatically in the developed RL environment. To handle the large state space, we use a Graph Neural Network (GNN) to compute a fixed-length feature vector that represents the information of vertiports and eVTOLs. The demand profile and passenger fare are modeled as time-series data and extracted by a Transformer encoder. The cost of per-passenger transportation is represented by a learned feature vector from a feedforward network. Corridor closure information $T^{\text{corr}}$ is flattened and concatenated with the aforementioned embeddings.

  \textbf{Action Space:}
During each decision-making instance, there will be one eVTOL (or aircraft) that will be deciding its next destination, i.e., select an action, $V^{\text{end}}_{k,l}$ $\in$ $V_{v}$), where $V_{v}$ includes all available vertiports. If it chooses the vertiport where it is currently located, it will wait or idle for 15 mins before triggering a new decision instance. 
We also consider a masking mechanism to prevent the selection of infeasible vertiport journeys, i.e., ones that violate any of the constraints presented in Eqs. \ref{c1}, \ref{c2}, and \ref{c3}.

\textbf{Reward:} We consider a delayed reward strategy, where the total reward computed at the end of the episode is the ratio of the profit (Eq. \ref{eq:objective_function}) to the maximum possible episodic profit, i.e.,
$\sum_{i \in V,j \in V,t \in [T^{\text{start}}, T^{\text{end}}]} (Q(i,j,t) \times F^{\text{passenger}}_{i,j,t})$. 

  \textbf{State Transition:} Uncertain route closure and take-off delay, and uncertain variations in demand lead to a stochastic state transition here, which is computed by the simulation. 

\vspace{-.1cm}
\section{Policy Architecture \& Learning Framework}
\label{sec:Proposed_solution}
  % In this work, we develop a learning framework, which will be used to assign actions to the eVTOLs during a decision-making instance. The policy network takes in the state information during each time step and outputs an action for each eVTOL. The policy network consists (shown in figure \ref{fig:policy}) of GNN-based encoders, Transformer network-based encoders, feedforward network, and Multi-head attention (MHA) based decoder. The state information is encoded as learnable fixed-length feature vectors (of length $l_{\text{embed}}$) by the encoder and a context module and will be used by the decoder to compute the action sequentially. Further information regarding the proposed state encoding and action decoding is discussed in the following section.
We develop an RL framework to compute actions for individual eVTOLs during decision-making instances. The policy model, as illustrated in Fig. \ref{fig:policy}, is called when a fully charged eVTOL is waiting to be assigned a destination. If an eVTOL remains grounded due to route closures, destination crowding, or insufficient battery, the policy model is re-queried after a 15-min wait. The policy model takes the state as input and outputs the probability of selecting the next destination for the eVTOL. As shown in Fig. \ref{fig:policy}), it combines GNN-based and Transformer-based encoders, a feedforward network, and a Multi-head Attention (MHA) based decoder. The state information is encoded as fixed-length (of $l_{\text{embed}}$)  feature vectors using encoders and a context module. These vectors are used by the decoder to compute actions sequentially. We call the proposed policy network as \textbf{Cap}sule \textbf{T}ransformer \textbf{A}ttention-mechanism \textbf{I}ntegrated \textbf{N}etwork (\textbf{CapTAIN}). Each component of this policy model is further explained below.

\vspace{-.1cm}
\subsection{State Encoding}  

  The state information consists of the information of UAM vertiports, eVTOLs, passenger demand, passenger fare, passenger transportation costs, and electricity pricing. This section describes how the state information is encoded.
\vspace{-.1cm}

\subsubsection{Vertiport and eVTOL State Encoding with GNN:} \label{Sec:GCAPCN} 
\begin{figure}[!t]
    \centering
    \includegraphics[width=0.99\columnwidth]{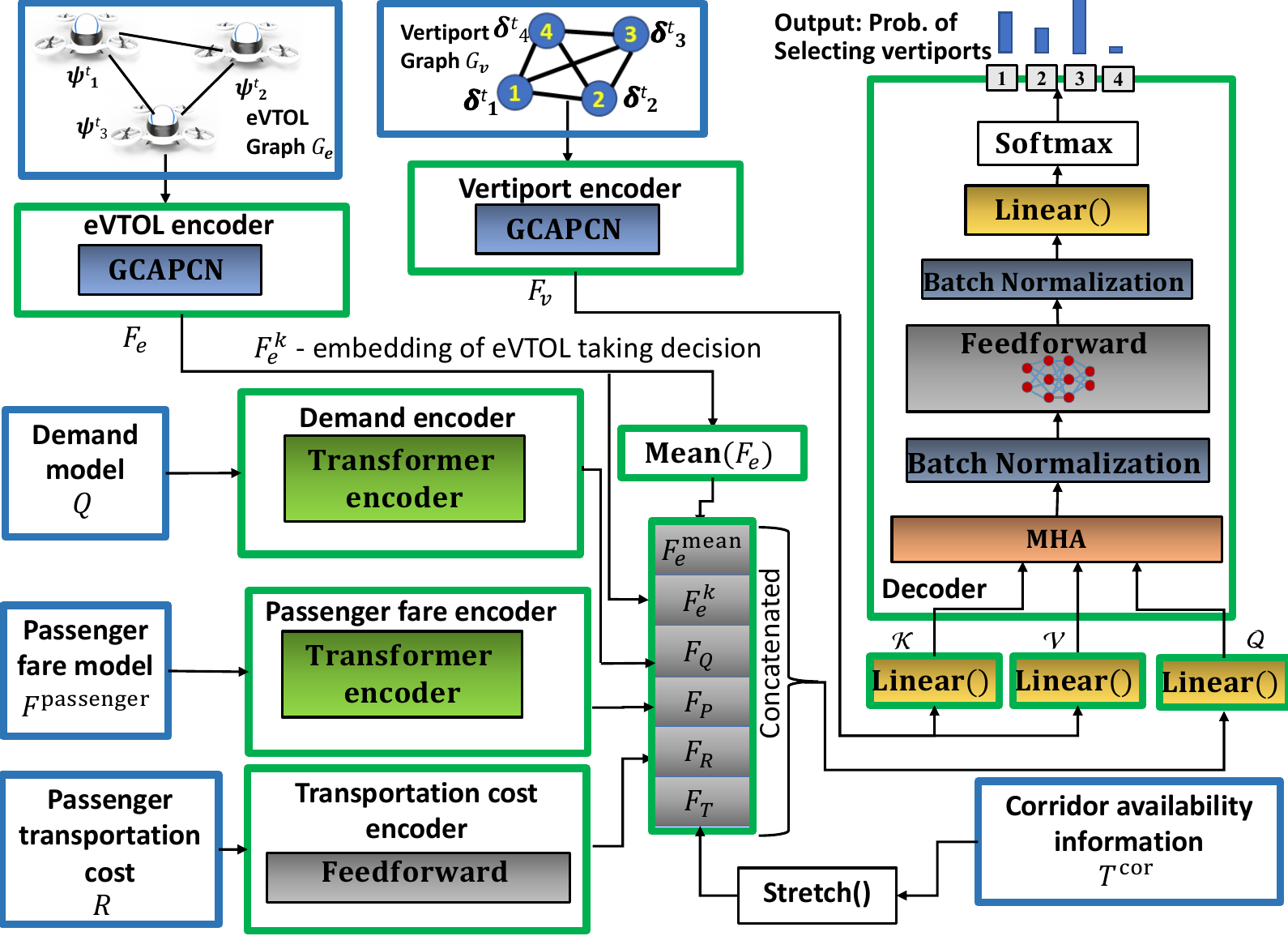}
    \caption{
The CapTAIN policy network consists of GCAPCN and Transformer encoders, and a decoder. Green and blue blocks represent the policy and the state space, respectively.}
    \label{fig:policy}
\end{figure}

We use a GNN to compute feature vectors for the vertiport and eVTOL state information, both of which are abstracted as graphs. Specifically, we employ a Graph Capsule Convolutional Neural Network (GCAPCN) for learning local and global structures with node properties (Fig. \ref{fig:policy} top-left). The node embedding in GCAPCN is permutation invariant, similar to the approach described in \cite{paul2022graph}. GCAPCN is a GNN introduced in \cite{Verma2018} to address the limitations of Graph Convolutional Neural Networks (GCN), enabling the encoding of global information using capsule networks as presented in \cite{Hinton}. Further description of the GCAPCN architecture can be found in \cite{paul2022graph}. We use GCAPCN to
compute node embeddings for the vertiport and eVTOL graphs, $F_{e} \in \mathbb{R}^{N_{K} \times l_{\text{embed}}}$ and $F_{v}\in \mathbb{R}^{N \times l_{\text{embed}}}$ respectively, where $l_{\text{embed}}$ is the embedding length. 
  \vspace{-.1cm}

\subsubsection{Transformer-based Encoding of Demand \& Passenger Fare: }\label{subsubsec:transformer_timeseries} Here, we utilize a Transformer Architecture to compute learnable embeddings for entities that can be represented as time-series data, such as demand distribution over time and passenger fare (Fig. \ref{fig:policy} bottom-left). The Transformer architecture, originally introduced in \cite{VaswaniSPUJGKP17} and widely used in applications such as Natural Language Processing \cite{VaswaniSPUJGKP17} and time series forecasting \cite{zhou2021informer}, follows an encoder-decoder approach based on self-attention. The encoder maps an input sequence to continuous representations, and the decoder generates an output sequence by attending to relevant information. In our work, we utilize the Transformer encoder to compute continuous and learnable feature vectors for time-series data.
%\subsubsection{Demand distribution and passenger fare encoding}\label{sec:demand_encoding}
%
The forecasted demand between vertiports during each hour ($Q$) and the passenger fare ($F^{\text{passenger}}$) are processed by two separate Transformer networks into learned feature vectors, $F_{Q}\in \mathbb{R}^{l_{\text{embed}}}$ and $F_{P}\in \mathbb{R}^{l_{\text{embed}}}$, respectively.
\vspace{-.1cm}

\subsubsection{Passenger transportation cost encoding:} \label{sec:passenger_transportation_cost_encoding}
The passenger transportation information $(R)$ can be considered as a matrix of size $N \times N$. This information can be encoded as a feature vector $F_{R}$ of length $l_{\text{embed}}$ by passing through a feedforward (FF) network.
\vspace{-.1cm}

\subsubsection{Corridor Availability encoding:}\label{subsubsec:corridor_availability}
For every corridor, we keep track of the time at which it is safe for a new eVTOL to enter the corridor (Fig. \ref{fig:decision-making}) subject to minimum separation requirements governed by safety. The corridor availability $T^{\text{cor}}$, a tensor of size $N \times N \times 2$. $T^{\text{cor}}$ is stretched into a 1-D vector $F_{T}$ of length $N \times N \times 2$, as shown in Fig. \ref{fig:policy} bottom-right.
\vspace{-.1cm}

\subsubsection{Context:} The context vector is computed by taking the linear transformation of a concatenated vector of the mean of the eVTOL node embedding ($F_{e}$) given by the GCAPCN encoder, the embedding of the eVTOL taking decision ($F_{e}^{k}$), demand and passenger fare encoding given by the transformers ($F_{Q}$ and $F_{P}$), the transportation cost encoding ($F_{R}$) given by the FF, and the corridor availability vector ($F_{T}$). 
\vspace{-.1cm}

\subsection{Action Decoding}
We use a Multi-head Attention-based decoder as shown in Fig. \ref{fig:policy} top-right, to compute which vertiport to visit next, given the vertiport node embeddings ($F_{v}$) and the Context vector. 
The attention mechanism computes compatibility scores between the context and the node embeddings, selecting the destination for the eVTOL based on the highest compatibility score. The choice of the current vertiport itself indicates a stay-idle decision. 
% \subsection{Training Algorithm} \label{subsec:Training_Algorithm}\vspace{-.2cm}
%%%%%%%%%%%%%%%%%%%%%%%%%%%%%%%%%%%%%%%%%%

\vspace{-.1cm}
\subsection{Simulation Environment \& Training Algorithm} \label{sec:sim}
%%%%%%%%%%%%%%%%%%%%%%%%%%%%%%%%%%%%%%%%%%
We considered a hypothetical area of 50 x 50 sq miles with 8 vertiport locations (Fig. \ref{fig:decision-making}), including 2 vertistops, $V_{s}$ (1 and 6). Four vertiports (0, 4, 6 and 7) are designated as high-traffic ($V_{B}$) due to their high trip demand. The location of vertiports remained the same for each training scenario, while the hourly demand values changed for each episode. The simulation environment is implemented in Python using the OpenAI Gym interface, making it compatible with standard RL training algorithms such as A2C \cite{Konda2003OnActorCriticA} and PPO \cite{schulman2017proximal} through the stable-baselines3 library. To train the policy network, here we use PPO from \textbf{stable-baselines3} \cite{stable-baselines3}. A 2-layer neural network with input and intermediate length $l_{\text{embed}}$ and LeakyReLU activation is used as the value network.
\vspace{-.1cm}

\section{Experimental Evaluation}\label{sec:experimental_evaluation}
% \begin{itemize}
%     \item Training convergence and training time
%     \item Testing for generalizability
%     \item Ablation study
%     \item Analysis of total delay
%     \item if results are close, then do statistical analysis
%     \item Total trips made
%     \item When the probaiblity of route closure is very high. Will the number wait action be more? If yes, it means its a conservative approach.
% \end{itemize}
In order to assess the importance of each principal component of the proposed CapTAIN policy model, e.g., the novel encoder choices, we train two alternate policy models, using two GPUs (NVIDIA Tesla V100, 16GB RAM). The first of these alternate models ablates the two GCAPCN encoders by replacing it with a Multi-Layered Perceptron (MLP) and is named the \textbf{Ablation-GCAPCN}, while the second one ablates the Transformer network by replacing it with an MLP, and is named the \textbf{Ablation-Transformer}. These alternate models are also compared against two other baselines (Section \ref{subsec:generalizability}). 

The first baseline is given by a \textbf{standard elitist Genetic Algorithm (GA)} that uses a population size of 30, max iteration of 30, mutation probability of 0.1, elite ratio of 0.1, and crossover probability of 0.5. Here, the GA is implemented in batches of 30 sequential decisions. During each batch, the optimization variables are the 30 decisions where each decision takes an integer value between 1 and N (for each vertiport), and the objective function is computed using Eq. \ref{eq:objective_function} at the end of the 30 decisions. We take the 30 decision variables and run the simulation with these 30 decisions implemented sequentially resulting in an updated environment state. The optimization of the next 30 decisions starts with the current state of the environment. This is continued until the end of an episode. A standard Python package\footnote{\url{https://pypi.org/project/geneticalgorithm/}} is used to implement the GA.

The second baseline method is a \textbf{Feasibility Preserving Random Walk (Feas-RND)} approach that randomly selects a vertiport from the set of feasible choices (satisfying Eqs. \ref{c1},\ref{c2}, \ref{c3}, and \ref{c4}) as the next destination during each decision-making instance.

Here onward, the Ablation-GCAPCN and Ablation-Transformer policy models are called the two learning-based baselines, while the GA and Feas-RND are called the two non-learning-based baselines.
\vspace{-.1cm}
\subsection{Training \& Convergence}\label{subsec:training_convergence}
\begin{figure}\hspace{-.8cm}
     \begin{minipage}[t]{0.6\linewidth}
        % \centering
%        \vspace{.5cm}
        \includegraphics[width=\linewidth]{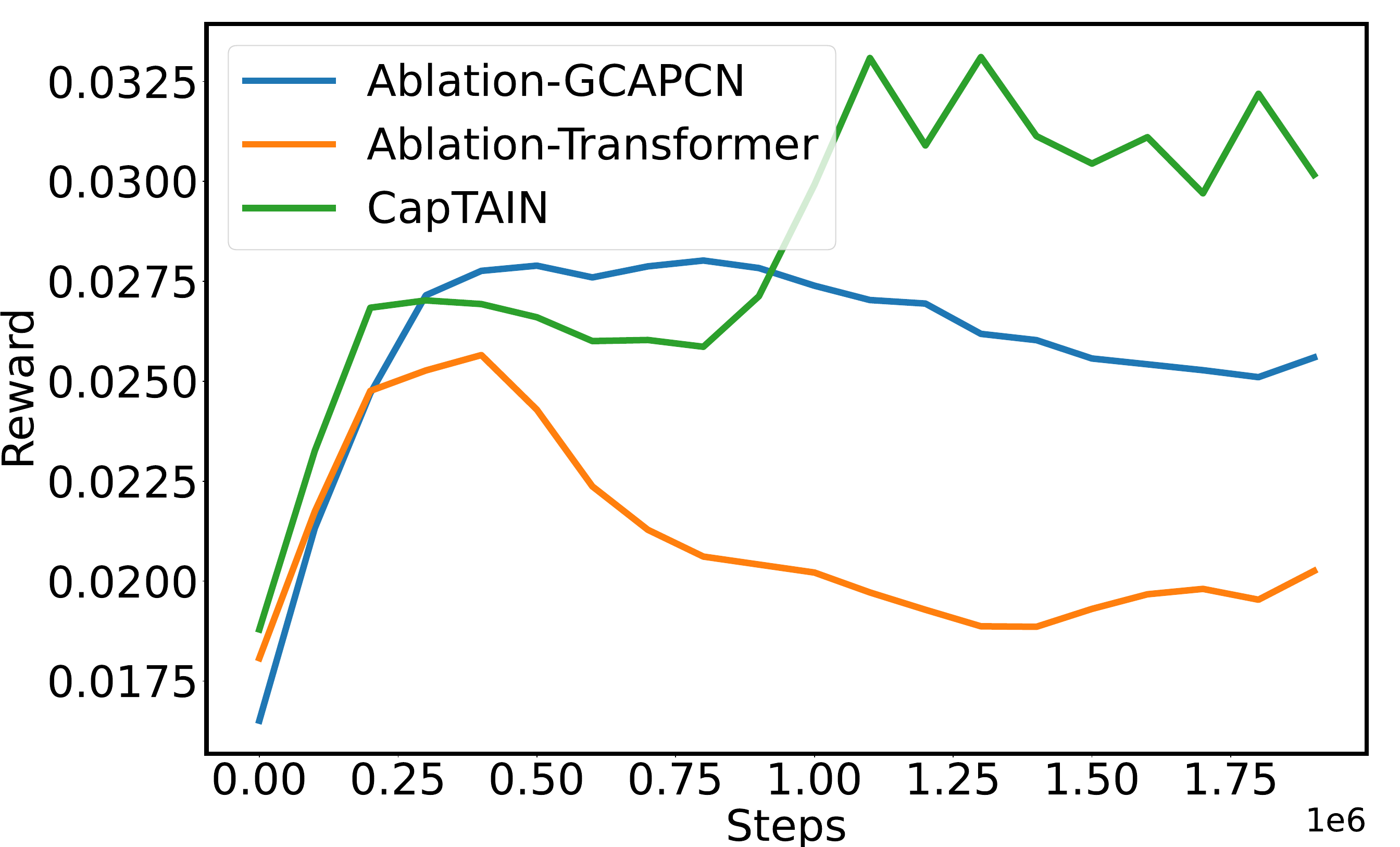}
        \caption{Convergence plot for all the trained models}
        \label{fig:convergence} 
    \end{minipage} \hspace{.08cm}
    \begin{minipage}[t]{0.30\linewidth}
    % \vspace{-0.1cm}
        \label{table:training_algo_info}
%        \vspace{-.5cm}
        % \centering
        \begin{small}
        \begin{sc}
        \vspace{-2.5cm}
        \begin{tabular}{|l|c|}
        \toprule
         Details &  Values \\ 
        \midrule
        \textit{Total steps}   & 2e6  \\
        \textit{Rollout size}   & 20000   \\
        \textit{Optimizer}   & \textit{Adam}  \\
        \textit{Learning step} & 1e-7 \\
        \textit{Entropy coef} & 0.01  \\
        \textit{Value coef} & 0.5  \\
        \bottomrule
        \end{tabular}
        \captionof{table}{Training settings}
        \end{sc}
        \end{small}
    \end{minipage}
    \vspace{-.6cm}
\end{figure}
% \begin{table}[!ht]
% \begin{table}[h]
% \caption{Training settings}
% \label{table:training_algo_info}
% \begin{center}
% \begin{small}
% \begin{sc}
% \begin{tabular}{lc}
% \toprule
%  Details &  Values \\ 
% \midrule
% \textit{Total steps}   & 2e6  \\
% \textit{Rollout size}   & 20000   \\
% \textit{Optimizer}   & \textit{Adam}  \\
% \textit{Learning step} & 1e-7 \\
% \textit{Entropy coef} & 0.01  \\
% \textit{Value coef} & 0.5  \\
% \bottomrule
% \end{tabular}
% \end{sc}
% \end{small}
% \end{center}
% \end{table}
The three policies are trained using PPO for 2 million steps based on the parameters in Table 1. From the convergence plots in Fig. \ref{fig:convergence}, it can be seen that CapTAIN converges to a higher episodic mean reward compared to the other two policies. Ablation-Transformer episodic rewards improved until 900K steps, after which the performance started to deteriorate. 
This demonstrates the utility of the combination of the two encoder components (GCAPCN and Transformer) to provide better learning capability. Next, these trained models are tested against the non-learning-based methods (GA and Feas-RND), with the latter implemented on a 2.6 GHz Intel core i7 MacOS 11.2.3 system.

\vspace{-.1cm}

\subsection{Generalizability \& Ablation Study}\label{subsec:generalizability}
\begin{figure}[!htb]
\vspace{-.3cm}
    \centering
    \includegraphics[width=\columnwidth, height=.5\columnwidth]{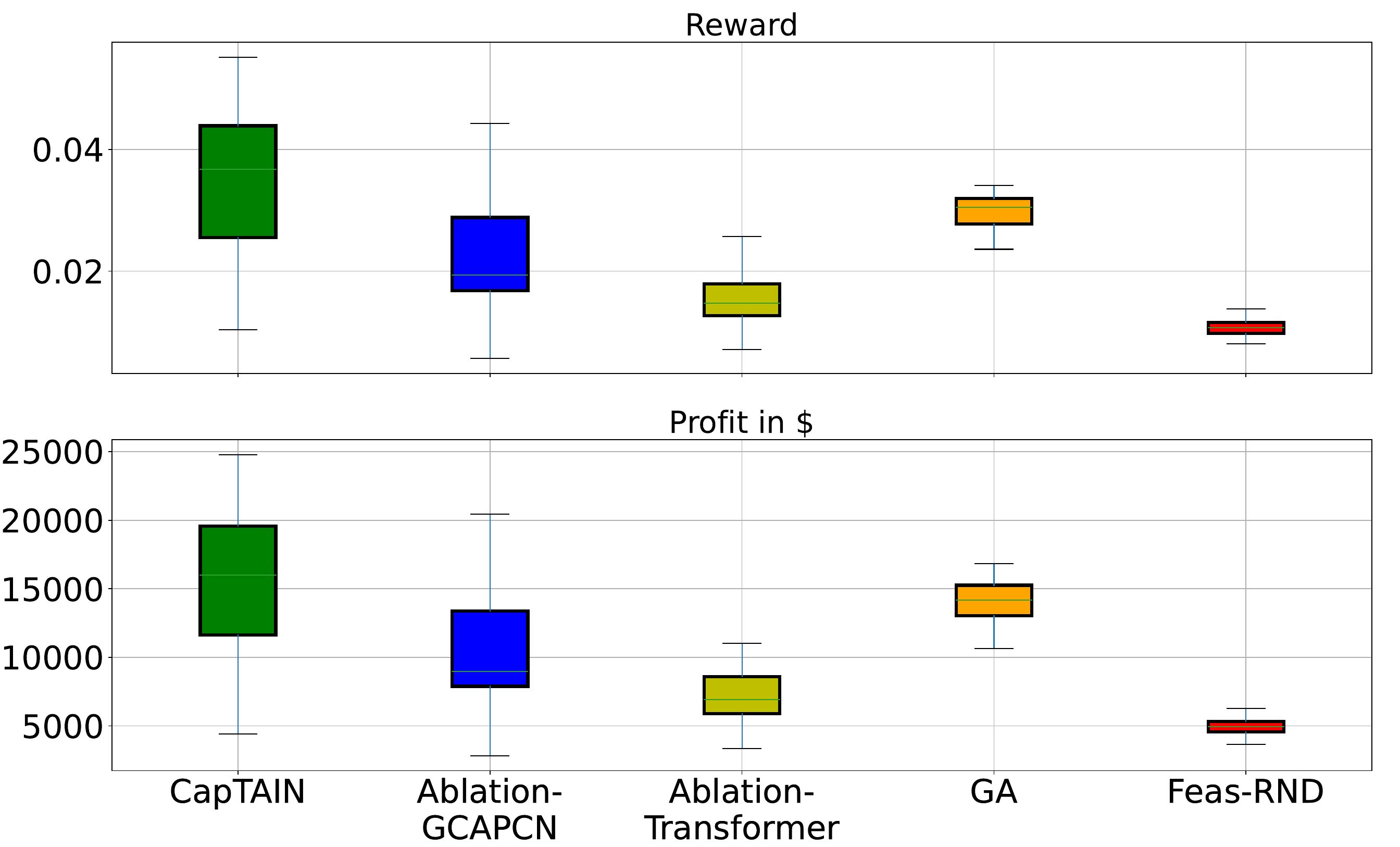}\vspace{-.3cm}
    \caption{Comparative analysis of the 5 methods on total episodic reward, total episodic profit, total idle decisions, and total flight decisions.}
    \label{fig:generalizability} 

\end{figure}

The trained RL models and baselines are tested on 100 unseen episodes, which are each defined as a 12 hr (6.00 am--6.00 pm) operation of the UAM network. Figure \ref{fig:generalizability} compares the results across all 100 scenarios based on the mean episodic reward and episodic profit. Figure \ref{fig:secondary} provides further comparisons in terms of the following metrics: total number stay-idle decisions, total number of flight decisions, total idle time, total flight time and total delay, computed across all eVTOLs over entire training episodes.
It is observed from Fig. \ref{fig:generalizability} that CapTAIN outperforms GA in terms of the average episodic rewards, achieving a mean reward of 0.34 vs. the 0.29 mean reward obtained by GA. To assess the significance of this difference, we perform a statistical T-test, with the null hypothesis being that both methods' mean performance is the same. The $p\text{-value}$ is found to be 6$\times$$10^{-6} (<.05)$, which indicates that CapTAIN has a statistically significant performance advantage over GA. As expected, Feas-RND performed the worst.

As seen from Fig. \ref{fig:generalizability}, the comparative trend (or ranking in terms) of the total episodic profit is very similar to that the episodic reward across the five methods tested, with CapTAIN performing the best. It is also notable that the Ablation-Transformer performs worse than Ablation-GCAPCN. This shows that the Transformer encoder contributes more strongly to generalizability compared to GCAPCN, at least under the current problem settings.

\vspace{-.1cm}

\subsubsection{Computation Time analysis}\label{subsubsec:computation_time}
The solution computing time for the GA is determined by adding the total time for which the optimization (in batches) is performed. Similarly for CapTAIN and the other learning-based methods, the total episodic computing time is the sum of the computing times for the forward propagation through the policy network throughout an episode. It's found that the average episodic computation time required by the GA is about 1,774 seconds, while for CapTAIN, it is only 2.1 seconds.

\vspace{-.1cm}

\subsection{Further Analysis of Decision-Making}\label{subsec:Qualitative analysis}
\begin{figure}[!htb]
    \centering
    \includegraphics[width=\columnwidth, height=1.38\columnwidth]{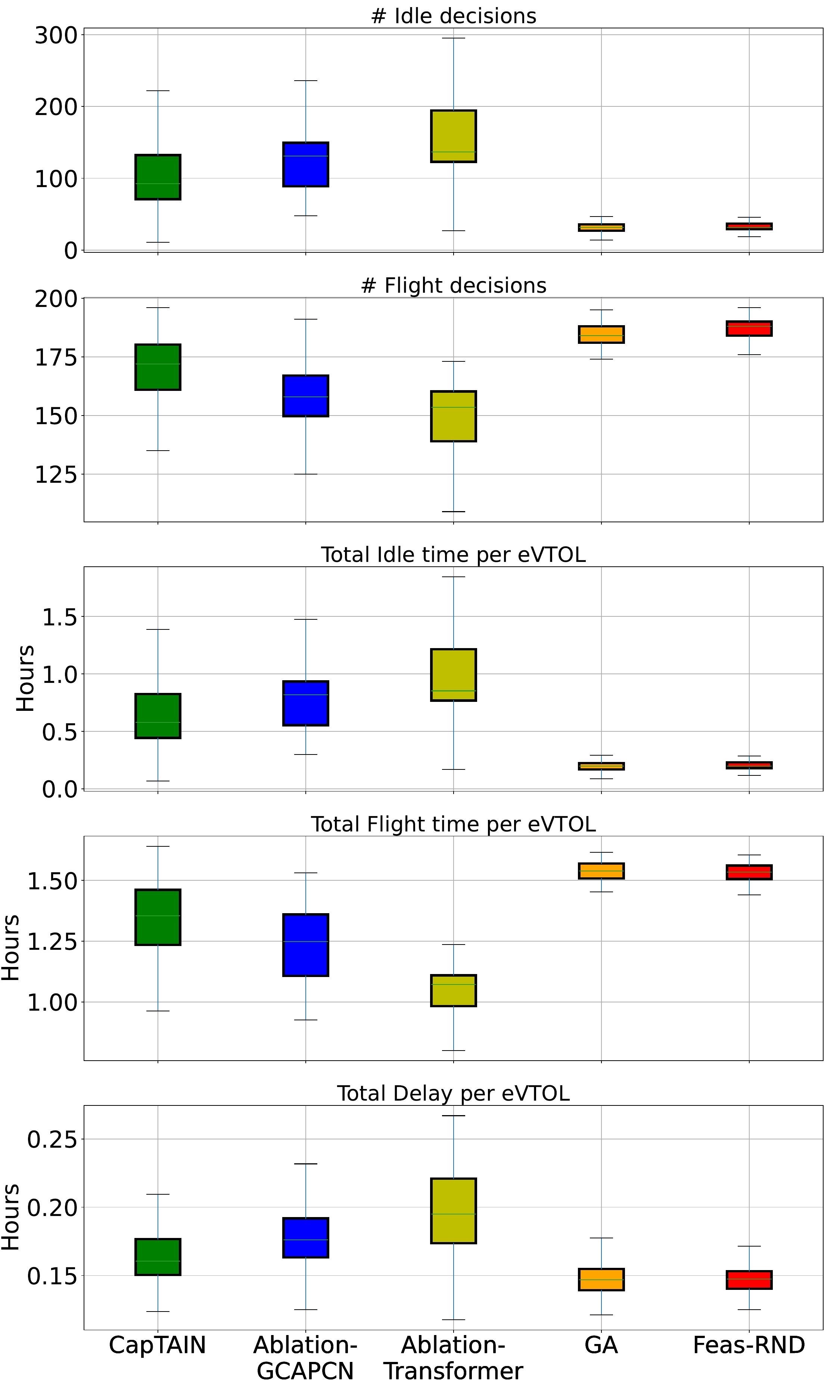}\vspace{-.5cm}
    \caption{
Comparing the effect of the 5 methods on total idle time, total flight time, and total delay time per eVTOL.}
    \label{fig:secondary}
\end{figure}

In order to physically interpret the diverse nature of the decision-making by the different methods, we track the total number of idle decisions, the total number of flight decisions, and the total episodic flight time, idle time and delays per eVTOL. From Fig. \ref{fig:secondary}, it can be seen that Feas-RND and GA have comparatively more number of flight decisions and fewer idle decisions, and consequently higher episodic flight time per eVTOL and lower episodic idle time per eVTOL (Fig. \ref{fig:secondary}), compared to CapTAIN. Interestingly, these observations show that more flights do not necessarily result in greater overall profits. This phenomenon is caused by the difference in demand and pricing across peak (higher fares) and off-peak (lower fares) hours. Here CapTAIN is taking advantage of this difference to provide better trade-offs in a number of peak/off-peak flights, leading to more favorable profit generation. 
Both GA and Feas-RND are also observed from Fig. \ref{fig:secondary} to result in lower total episodic delay per eVTOL, compared to CapTAIN. This shows that CapTAIN compromises delays with the objective of a higher profit, which could also be an artifact of delays not directly affecting demand or fares. %Future modification of the objective or reward function to explicitly include aggregated delays as a penalty term could potentially improve the realism of the simulations and resulting policies.
\vspace{-.4cm}

\vspace{-.6cm}
\section{CONCLUSIONS} \label{sec:conclusion}
\vspace{-.2cm}
We proposed a graph RL approach using a new encoder-decoder policy architecture to perform UAM fleet scheduling, one that uniquely considers uncertainties (due to delays, aircraft downtime, and route closures), time-varying demand, vertiport constraints, and airspace constraints. The state of eVTOL aircraft and vertiport occupancy were expressed as graphs, while demand forecast and fares were treated as time-series data. Our novel architecture (CapTAIN) comprises Graph Capsule Convolutional Networks, Transformer encoders, feedforward context layers, and a Multi-head Attention-based decoder to compute sequential actions. Each of these components of CapTAIN is designed to play a specific role that caters to a specific complexity of the fleet scheduling problem, such as generalizable embedding of the structural information of the vertiports and eVTOL graphs, and transforming time-series data for demand and passenger fare into context vectors. The policy was trained using PPO and evaluated on 100 unseen scenarios. Compared to non-learning-based methods (GA and Feas-RND), our CapTAIN achieved better performance with up to 3 orders of magnitude faster computation time than GA. Ablation studies showed that the Transformer encoder had a greater impact on performance than graph neural net encoders. Future directions include extending planning horizons and incorporating end-of-episode constraints to alleviate the limitations of the current myopic training implementation of the policy. Decentralized policies suitable for more realistic scenarios with multiple stakeholders operating vertiports and eVTOL fleets can also be explored in the future.
\vspace{-.3cm}

\section*{Acknowledgments}\vspace{-.2cm}
This work was supported by the Office of Naval Research (ONR) award N00014-21-1-2530 and the National Science Foundation (NSF) award CMMI 2048020. Any opinions, findings, conclusions, or recommendations expressed in this paper are those of the authors and do not necessarily reflect the views of the ONR or the NSF.
\vspace{-.3cm}

% \appendix
% %Appendix A

% \begin{acks}
%   The authors would like to thank Dr. Yuhua Li for providing the
%   matlab code of  the \textit{BEPS} method. 

%   The authors would also like to thank the anonymous referees for
%   their valuable comments and helpful suggestions. The work is
%   supported by the \grantsponsor{GS501100001809}{National Natural
%     Science Foundation of
%     China}{http://dx.doi.org/10.13039/501100001809} under Grant
%   No.:~\grantnum{GS501100001809}{61273304}
%   and~\grantnum[http://www.nnsf.cn/youngscientsts]{GS501100001809}{Young
%     Scientsts' Support Program}.

% \end{acks}

\bibliographystyle{ACM-Reference-Format}
\bibliography{irmas.bib} 


\begin{thebibliography}{41}

%%% the Web address.

\ifx \showCODEN    \undefined \def \showCODEN     #1{\unskip}     \fi
\ifx \showDOI      \undefined \def \showDOI       #1{#1}\fi
\ifx \showISBNx    \undefined \def \showISBNx     #1{\unskip}     \fi
\ifx \showISBNxiii \undefined \def \showISBNxiii  #1{\unskip}     \fi
\ifx \showISSN     \undefined \def \showISSN      #1{\unskip}     \fi
\ifx \showLCCN     \undefined \def \showLCCN      #1{\unskip}     \fi
\ifx \shownote     \undefined \def \shownote      #1{#1}          \fi
\ifx \showarticletitle \undefined \def \showarticletitle #1{#1}   \fi
\ifx \showURL      \undefined \def \showURL       {\relax}        \fi
% invisible to TeX
\providecommand\bibfield[2]{#2}
\providecommand\bibinfo[2]{#2}
\providecommand\natexlab[1]{#1}
\providecommand\showeprint[2][]{arXiv:#2}

\bibitem[ele({[n.\,d.]})]%
        {electric_price}
 \bibinfo{year}{[n.\,d.]}\natexlab{}.
\newblock \bibinfo{title}{See electric rates available to your home/business (updated today):}.
\newblock
\newblock
\urldef\tempurl%
\url{https://www.electricchoice.com/electricity-prices-by-state/}
\showURL{%
\tempurl}


\bibitem[FAA(2023)]%
        {FAA_CON_OPS}
 \bibinfo{year}{2023}\natexlab{}.
\newblock \bibinfo{title}{Urban Air Mobility (UAM) Concept of Operations -v2.0}.
\newblock
\newblock
\urldef\tempurl%
\url{https://www.faa.gov/sites/faa.gov/files/Urban\%20Air\%20Mobility\%20\%28UAM\%29\%20Concept\%20of\%20Operations\%202.0\_0.pdf}
\showURL{%
\tempurl}


\bibitem[Barrett et~al\mbox{.}(2019)]%
        {barrett2019exploratory}
\bibfield{author}{\bibinfo{person}{Thomas~D Barrett}, \bibinfo{person}{William~R Clements}, \bibinfo{person}{Jakob~N Foerster}, {and} \bibinfo{person}{Alex~I Lvovsky}.} \bibinfo{year}{2019}\natexlab{}.
\newblock \showarticletitle{Exploratory combinatorial optimization with reinforcement learning}.
\newblock \bibinfo{journal}{\emph{arXiv preprint arXiv:1909.04063}} (\bibinfo{year}{2019}).
\newblock


\bibitem[Chuwang and Chen(2022)]%
        {chuwang2022forecasting}
\bibfield{author}{\bibinfo{person}{Dung~David Chuwang} {and} \bibinfo{person}{Weiya Chen}.} \bibinfo{year}{2022}\natexlab{}.
\newblock \showarticletitle{Forecasting daily and weekly passenger demand for urban rail transit stations based on a time series model approach}.
\newblock \bibinfo{journal}{\emph{Forecasting}} \bibinfo{volume}{4}, \bibinfo{number}{4} (\bibinfo{year}{2022}), \bibinfo{pages}{904--924}.
\newblock


\bibitem[Fernando et~al\mbox{.}(2023)]%
        {fernando2023graph}
\bibfield{author}{\bibinfo{person}{Malintha Fernando}, \bibinfo{person}{Ransalu Senanayake}, \bibinfo{person}{Heeyoul Choi}, {and} \bibinfo{person}{Martin Swany}.} \bibinfo{year}{2023}\natexlab{}.
\newblock \showarticletitle{Graph Attention Multi-Agent Fleet Autonomy for Advanced Air Mobility}.
\newblock \bibinfo{journal}{\emph{arXiv preprint arXiv:2302.07337}} (\bibinfo{year}{2023}).
\newblock


\bibitem[Hinton et~al\mbox{.}(2011)]%
        {Hinton}
\bibfield{author}{\bibinfo{person}{Geoffrey~E. Hinton}, \bibinfo{person}{Alex Krizhevsky}, {and} \bibinfo{person}{Sida~D. Wang}.} \bibinfo{year}{2011}\natexlab{}.
\newblock \showarticletitle{Transforming Auto-Encoders}. In \bibinfo{booktitle}{\emph{Artificial Neural Networks and Machine Learning -- ICANN 2011}}, \bibfield{editor}{\bibinfo{person}{Timo Honkela}, \bibinfo{person}{W{\l}odzis{\l}aw Duch}, \bibinfo{person}{Mark Girolami}, {and} \bibinfo{person}{Samuel Kaski}} (Eds.). \bibinfo{publisher}{Springer Berlin Heidelberg}, \bibinfo{address}{Berlin, Heidelberg}, \bibinfo{pages}{44--51}.
\newblock
\showISBNx{978-3-642-21735-7}


\bibitem[Hwang and Cheng(2001)]%
        {Hwang2001}
\bibfield{author}{\bibinfo{person}{Shyh~In Hwang} {and} \bibinfo{person}{Sheng~Tzong Cheng}.} \bibinfo{year}{2001}\natexlab{}.
\newblock \showarticletitle{{Combinatorial Optimization in Real-Time Scheduling: Theory and Algorithms}}.
\newblock \bibinfo{journal}{\emph{Journal of Combinatorial Optimization}} (\bibinfo{year}{2001}).
\newblock
\showISSN{13826905}
\urldef\tempurl%
\url{https://doi.org/10.1023/A:1011449311477}
\showDOI{\tempurl}


\bibitem[III({[n.\,d.]})]%
        {iii}
\bibfield{author}{\bibinfo{person}{Woodrow~Bellamy III}.} \bibinfo{year}{[n.\,d.]}\natexlab{}.
\newblock \bibinfo{title}{Evtol investments will continue billion dollar trend in 2021}.
\newblock
\newblock
\urldef\tempurl%
\url{http://interactive.aviationtoday.com/avionicsmagazine/february-march-2021/evtol-investments-will-continue-billion-dollar-trend-in-2021/}
\showURL{%
\tempurl}


\bibitem[Jacob et~al\mbox{.}(2022)]%
        {9750805}
\bibfield{author}{\bibinfo{person}{Roshni~Anna Jacob}, \bibinfo{person}{Steve Paul}, \bibinfo{person}{Wenyuan Li}, \bibinfo{person}{Souma Chowdhury}, \bibinfo{person}{Yulia~R. Gel}, {and} \bibinfo{person}{Jie Zhang}.} \bibinfo{year}{2022}\natexlab{}.
\newblock \showarticletitle{Reconfiguring Unbalanced Distribution Networks using Reinforcement Learning over Graphs}. In \bibinfo{booktitle}{\emph{2022 IEEE Texas Power and Energy Conference (TPEC)}}. \bibinfo{pages}{1--6}.
\newblock
\urldef\tempurl%
\url{https://doi.org/10.1109/TPEC54980.2022.9750805}
\showDOI{\tempurl}


\bibitem[Jonas et~al\mbox{.}({[n.\,d.]})]%
        {jonas_jonas}
\bibfield{author}{\bibinfo{person}{Adam Jonas}, \bibinfo{person}{Adam Jonas}, \bibinfo{person}{Head of Global~Auto}, {and} \bibinfo{person}{Shared~Mobility Research}.} \bibinfo{year}{[n.\,d.]}\natexlab{}.
\newblock \bibinfo{title}{Are flying cars preparing for takeoff?}
\newblock
\newblock
\urldef\tempurl%
\url{https://www.morganstanley.com/ideas/autonomous-aircraft}
\showURL{%
\tempurl}


\bibitem[Kaempfer and Wolf(2018)]%
        {Kaempfer2018LearningTM}
\bibfield{author}{\bibinfo{person}{Yoav Kaempfer} {and} \bibinfo{person}{Lior Wolf}.} \bibinfo{year}{2018}\natexlab{}.
\newblock \showarticletitle{Learning the Multiple Traveling Salesmen Problem with Permutation Invariant Pooling Networks}.
\newblock \bibinfo{journal}{\emph{ArXiv}}  \bibinfo{volume}{abs/1803.09621} (\bibinfo{year}{2018}).
\newblock


\bibitem[Kamra and Ayanian(2015)]%
        {kamra2015mixed}
\bibfield{author}{\bibinfo{person}{Nitin Kamra} {and} \bibinfo{person}{Nora Ayanian}.} \bibinfo{year}{2015}\natexlab{}.
\newblock \showarticletitle{A mixed integer programming model for timed deliveries in multirobot systems}. In \bibinfo{booktitle}{\emph{2015 IEEE International Conference on Automation Science and Engineering (CASE)}}. IEEE, \bibinfo{pages}{612--617}.
\newblock


\bibitem[Kim(2020)]%
        {8901431}
\bibfield{author}{\bibinfo{person}{Sang~Hyun Kim}.} \bibinfo{year}{2020}\natexlab{}.
\newblock \showarticletitle{Receding Horizon Scheduling of On-Demand Urban Air Mobility With Heterogeneous Fleet}.
\newblock \bibinfo{journal}{\emph{IEEE Trans. Aerospace Electron. Systems}} \bibinfo{volume}{56}, \bibinfo{number}{4} (\bibinfo{year}{2020}), \bibinfo{pages}{2751--2761}.
\newblock
\urldef\tempurl%
\url{https://doi.org/10.1109/TAES.2019.2953417}
\showDOI{\tempurl}


\bibitem[Konda and Tsitsiklis(2003)]%
        {Konda2003OnActorCriticA}
\bibfield{author}{\bibinfo{person}{V. Konda} {and} \bibinfo{person}{J. Tsitsiklis}.} \bibinfo{year}{2003}\natexlab{}.
\newblock \showarticletitle{OnActor-Critic Algorithms}.
\newblock \bibinfo{journal}{\emph{SIAM J. Control. Optim.}}  \bibinfo{volume}{42} (\bibinfo{year}{2003}), \bibinfo{pages}{1143--1166}.
\newblock


\bibitem[Kool et~al\mbox{.}(2019)]%
        {Kool2019}
\bibfield{author}{\bibinfo{person}{Wouter Kool}, \bibinfo{person}{Herke {Van Hoof}}, {and} \bibinfo{person}{Max Welling}.} \bibinfo{year}{2019}\natexlab{}.
\newblock \showarticletitle{{Attention, learn to solve routing problems!}}. In \bibinfo{booktitle}{\emph{7th International Conference on Learning Representations, ICLR 2019}}.
\newblock
\showeprint[arxiv]{1803.08475}


\bibitem[KrisshnaKumar et~al\mbox{.}(2023)]%
        {krisshnakumar2023fast}
\bibfield{author}{\bibinfo{person}{Prajit KrisshnaKumar}, \bibinfo{person}{Jhoel Witter}, \bibinfo{person}{Steve Paul}, \bibinfo{person}{Hanvit Cho}, \bibinfo{person}{Karthik Dantu}, {and} \bibinfo{person}{Souma Chowdhury}.} \bibinfo{year}{2023}\natexlab{}.
\newblock \showarticletitle{Fast Decision Support for Air Traffic Management at Urban Air Mobility Vertiports using Graph Learning}. In \bibinfo{booktitle}{\emph{2023 IEEE/RSJ International Conference on Intelligent Robots and Systems (IROS)}}. IEEE, \bibinfo{pages}{1580--1585}.
\newblock


\bibitem[Li et~al\mbox{.}(2018)]%
        {li2018combinatorial}
\bibfield{author}{\bibinfo{person}{Zhuwen Li}, \bibinfo{person}{Qifeng Chen}, {and} \bibinfo{person}{Vladlen Koltun}.} \bibinfo{year}{2018}\natexlab{}.
\newblock \showarticletitle{Combinatorial optimization with graph convolutional networks and guided tree search}. In \bibinfo{booktitle}{\emph{Advances in Neural Information Processing Systems}}. \bibinfo{pages}{539--548}.
\newblock


\bibitem[Miller et~al\mbox{.}(1960)]%
        {miller1960integer}
\bibfield{author}{\bibinfo{person}{Clair~E Miller}, \bibinfo{person}{Albert~W Tucker}, {and} \bibinfo{person}{Richard~A Zemlin}.} \bibinfo{year}{1960}\natexlab{}.
\newblock \showarticletitle{Integer programming formulation of traveling salesman problems}.
\newblock \bibinfo{journal}{\emph{Journal of the ACM (JACM)}} \bibinfo{volume}{7}, \bibinfo{number}{4} (\bibinfo{year}{1960}), \bibinfo{pages}{326--329}.
\newblock


\bibitem[M{\"u}hlenbein(1991)]%
        {gen-2}
\bibfield{author}{\bibinfo{person}{H. M{\"u}hlenbein}.} \bibinfo{year}{1991}\natexlab{}.
\newblock \showarticletitle{Parallel genetic algorithms, population genetics and combinatorial optimization}. In \bibinfo{booktitle}{\emph{Parallelism, Learning, Evolution}}, \bibfield{editor}{\bibinfo{person}{J.~D. Becker}, \bibinfo{person}{I.~Eisele}, {and} \bibinfo{person}{F.~W. M{\"u}ndemann}} (Eds.). \bibinfo{publisher}{Springer Berlin Heidelberg}, \bibinfo{address}{Berlin, Heidelberg}, \bibinfo{pages}{398--406}.
\newblock
\showISBNx{978-3-540-46663-5}


\bibitem[Nowak et~al\mbox{.}(2017)]%
        {nowak2017note}
\bibfield{author}{\bibinfo{person}{Alex Nowak}, \bibinfo{person}{Soledad Villar}, \bibinfo{person}{Afonso~S Bandeira}, {and} \bibinfo{person}{Joan Bruna}.} \bibinfo{year}{2017}\natexlab{}.
\newblock \showarticletitle{A note on learning algorithms for quadratic assignment with graph neural networks}.
\newblock \bibinfo{journal}{\emph{stat}}  \bibinfo{volume}{1050} (\bibinfo{year}{2017}), \bibinfo{pages}{22}.
\newblock


\bibitem[Paul and Chowdhury(2022a)]%
        {paul2022graph}
\bibfield{author}{\bibinfo{person}{Steve Paul} {and} \bibinfo{person}{Souma Chowdhury}.} \bibinfo{year}{2022}\natexlab{a}.
\newblock \showarticletitle{A Graph-based Reinforcement Learning Framework for Urban Air Mobility Fleet Scheduling}. In \bibinfo{booktitle}{\emph{AIAA AVIATION 2022 Forum}}. \bibinfo{pages}{3911}.
\newblock


\bibitem[Paul and Chowdhury(2022b)]%
        {paul2022scalable}
\bibfield{author}{\bibinfo{person}{Steve Paul} {and} \bibinfo{person}{Souma Chowdhury}.} \bibinfo{year}{2022}\natexlab{b}.
\newblock \showarticletitle{A Scalable Graph Learning Approach to Capacitated Vehicle Routing Problem Using Capsule Networks and Attention Mechanism}. In \bibinfo{booktitle}{\emph{International Design Engineering Technical Conferences and Computers and Information in Engineering Conference}}, Vol.~\bibinfo{volume}{86236}. American Society of Mechanical Engineers, \bibinfo{pages}{V03BT03A045}.
\newblock


\bibitem[Paul et~al\mbox{.}(2022)]%
        {capam_mrta}
\bibfield{author}{\bibinfo{person}{Steve Paul}, \bibinfo{person}{Payam Ghassemi}, {and} \bibinfo{person}{Souma Chowdhury}.} \bibinfo{year}{2022}\natexlab{}.
\newblock \showarticletitle{Learning Scalable Policies over Graphs for Multi-Robot Task Allocation using Capsule Attention Networks}. In \bibinfo{booktitle}{\emph{2022 International Conference on Robotics and Automation (ICRA)}}. IEEE, \bibinfo{pages}{8815--8822}.
\newblock


\bibitem[Peng et~al\mbox{.}(2022)]%
        {9925782}
\bibfield{author}{\bibinfo{person}{Xin Peng}, \bibinfo{person}{Vishwanath Bulusu}, {and} \bibinfo{person}{Raja Sengupta}.} \bibinfo{year}{2022}\natexlab{}.
\newblock \showarticletitle{Hierarchical Vertiport Network Design for On-Demand Multi-modal Urban Air Mobility}. In \bibinfo{booktitle}{\emph{2022 IEEE/AIAA 41st Digital Avionics Systems Conference (DASC)}}. \bibinfo{pages}{1--8}.
\newblock
\urldef\tempurl%
\url{https://doi.org/10.1109/DASC55683.2022.9925782}
\showDOI{\tempurl}


\bibitem[Peng et~al\mbox{.}(2021)]%
        {peng2021graph}
\bibfield{author}{\bibinfo{person}{Yun Peng}, \bibinfo{person}{Byron Choi}, {and} \bibinfo{person}{Jianliang Xu}.} \bibinfo{year}{2021}\natexlab{}.
\newblock \showarticletitle{Graph Learning for Combinatorial Optimization: A Survey of State-of-the-Art}.
\newblock \bibinfo{journal}{\emph{Data Science and Engineering}} (\bibinfo{year}{2021}), \bibinfo{pages}{1--23}.
\newblock


\bibitem[Pradeep and Wei(2018)]%
        {8569225}
\bibfield{author}{\bibinfo{person}{Priyank Pradeep} {and} \bibinfo{person}{Peng Wei}.} \bibinfo{year}{2018}\natexlab{}.
\newblock \showarticletitle{Heuristic Approach for Arrival Sequencing and Scheduling for eVTOL Aircraft in On-Demand Urban Air Mobility}. In \bibinfo{booktitle}{\emph{2018 IEEE/AIAA 37th Digital Avionics Systems Conference (DASC)}}. \bibinfo{pages}{1--7}.
\newblock
\urldef\tempurl%
\url{https://doi.org/10.1109/DASC.2018.8569225}
\showDOI{\tempurl}


\bibitem[Raffin et~al\mbox{.}(2021)]%
        {stable-baselines3}
\bibfield{author}{\bibinfo{person}{Antonin Raffin}, \bibinfo{person}{Ashley Hill}, \bibinfo{person}{Adam Gleave}, \bibinfo{person}{Anssi Kanervisto}, \bibinfo{person}{Maximilian Ernestus}, {and} \bibinfo{person}{Noah Dormann}.} \bibinfo{year}{2021}\natexlab{}.
\newblock \showarticletitle{Stable-Baselines3: Reliable Reinforcement Learning Implementations}.
\newblock \bibinfo{journal}{\emph{Journal of Machine Learning Research}} \bibinfo{volume}{22}, \bibinfo{number}{268} (\bibinfo{year}{2021}), \bibinfo{pages}{1--8}.
\newblock
\urldef\tempurl%
\url{http://jmlr.org/papers/v22/20-1364.html}
\showURL{%
\tempurl}


\bibitem[Rizzoli et~al\mbox{.}(2007)]%
        {Rizzoli2007}
\bibfield{author}{\bibinfo{person}{A~E Rizzoli}, \bibinfo{person}{R Montemanni}, \bibinfo{person}{E Lucibello}, {and} \bibinfo{person}{L~M Gambardella}.} \bibinfo{year}{2007}\natexlab{}.
\newblock \showarticletitle{{Ant colony optimization for real-world vehicle routing problems}}.
\newblock \bibinfo{journal}{\emph{Swarm Intelligence}} \bibinfo{volume}{1}, \bibinfo{number}{2} (\bibinfo{year}{2007}), \bibinfo{pages}{135--151}.
\newblock
\showISSN{1935-3820}
\urldef\tempurl%
\url{https://doi.org/10.1007/s11721-007-0005-x}
\showDOI{\tempurl}


\bibitem[Schoppmann(2022)]%
        {schoppmann2022operation}
\bibfield{author}{\bibinfo{person}{Marc~Josef Schoppmann}.} \bibinfo{year}{2022}\natexlab{}.
\newblock \emph{\bibinfo{title}{The operation of eVTOLs in the urban air mobility sector: use case \& operator assessment}}.
\newblock \bibinfo{thesistype}{Ph.\,D. Dissertation}.
\newblock


\bibitem[Schulman et~al\mbox{.}(2017)]%
        {schulman2017proximal}
\bibfield{author}{\bibinfo{person}{John Schulman}, \bibinfo{person}{Filip Wolski}, \bibinfo{person}{Prafulla Dhariwal}, \bibinfo{person}{Alec Radford}, {and} \bibinfo{person}{Oleg Klimov}.} \bibinfo{year}{2017}\natexlab{}.
\newblock \bibinfo{title}{Proximal Policy Optimization Algorithms}.
\newblock
\newblock
\showeprint[arxiv]{1707.06347}~[cs.LG]


\bibitem[Shihab et~al\mbox{.}(2020)]%
        {Shihab2020OptimalEF}
\bibfield{author}{\bibinfo{person}{Syed Arbab~Mohd Shihab}, \bibinfo{person}{Peng Wei}, \bibinfo{person}{Jie Shi}, {and} \bibinfo{person}{Nanpeng Yu}.} \bibinfo{year}{2020}\natexlab{}.
\newblock \showarticletitle{Optimal eVTOL Fleet Dispatch for Urban Air Mobility and Power Grid Services}.
\newblock \bibinfo{journal}{\emph{AIAA AVIATION 2020 FORUM}} (\bibinfo{year}{2020}).
\newblock


\bibitem[Sykora et~al\mbox{.}(2020)]%
        {Sykora2020}
\bibfield{author}{\bibinfo{person}{Quinlan Sykora}, \bibinfo{person}{Mengye Ren}, {and} \bibinfo{person}{Raquel Urtasun}.} \bibinfo{year}{2020}\natexlab{}.
\newblock \showarticletitle{{Multi-agent routing value iteration network}}. In \bibinfo{booktitle}{\emph{37th International Conference on Machine Learning, ICML 2020}}.
\newblock
\showISBNx{9781713821120}
\showeprint[arxiv]{2007.05096}


\bibitem[Thipphavong et~al\mbox{.}(2018)]%
        {thipphavong2018urban}
\bibfield{author}{\bibinfo{person}{David~P Thipphavong}, \bibinfo{person}{Rafael Apaza}, \bibinfo{person}{Bryan Barmore}, \bibinfo{person}{Vernol Battiste}, \bibinfo{person}{Barbara Burian}, \bibinfo{person}{Quang Dao}, \bibinfo{person}{Michael Feary}, \bibinfo{person}{Susie Go}, \bibinfo{person}{Kenneth~H Goodrich}, \bibinfo{person}{Jeffrey Homola}, {et~al\mbox{.}}} \bibinfo{year}{2018}\natexlab{}.
\newblock \showarticletitle{Urban air mobility airspace integration concepts and considerations}. In \bibinfo{booktitle}{\emph{2018 Aviation Technology, Integration, and Operations Conference}}. \bibinfo{pages}{3676}.
\newblock


\bibitem[Tolstaya et~al\mbox{.}(2020)]%
        {Tolstaya2020MultiRobotCA}
\bibfield{author}{\bibinfo{person}{Ekaterina~V. Tolstaya}, \bibinfo{person}{James Paulos}, \bibinfo{person}{Vijay~R. Kumar}, {and} \bibinfo{person}{Alejandro Ribeiro}.} \bibinfo{year}{2020}\natexlab{}.
\newblock \showarticletitle{Multi-Robot Coverage and Exploration using Spatial Graph Neural Networks}.
\newblock \bibinfo{journal}{\emph{ArXiv}}  \bibinfo{volume}{abs/2011.01119} (\bibinfo{year}{2020}).
\newblock


\bibitem[{United States. Federal Highway Administration}(2010)]%
        {Administration}
\bibfield{editor}{\bibinfo{person}{{United States. Federal Highway Administration}}} (Ed.). \bibinfo{year}{2010}\natexlab{}.
\newblock \showarticletitle{Our {Nation}'s {Highways} 2010}.
\newblock  \bibinfo{number}{FHWA-PL-10-023} (\bibinfo{date}{Jan.} \bibinfo{year}{2010}).
\newblock
\urldef\tempurl%
\url{https://rosap.ntl.bts.gov/view/dot/904}
\showURL{%
\tempurl}


\bibitem[Vaswani et~al\mbox{.}(2017)]%
        {VaswaniSPUJGKP17}
\bibfield{author}{\bibinfo{person}{Ashish Vaswani}, \bibinfo{person}{Noam Shazeer}, \bibinfo{person}{Niki Parmar}, \bibinfo{person}{Jakob Uszkoreit}, \bibinfo{person}{Llion Jones}, \bibinfo{person}{Aidan~N. Gomez}, \bibinfo{person}{Lukasz Kaiser}, {and} \bibinfo{person}{Illia Polosukhin}.} \bibinfo{year}{2017}\natexlab{}.
\newblock \showarticletitle{Attention Is All You Need}.
\newblock \bibinfo{journal}{\emph{CoRR}}  \bibinfo{volume}{abs/1706.03762} (\bibinfo{year}{2017}).
\newblock
\showeprint[arxiv]{1706.03762}
\urldef\tempurl%
\url{http://arxiv.org/abs/1706.03762}
\showURL{%
\tempurl}


\bibitem[Verma and Zhang(2018)]%
        {Verma2018}
\bibfield{author}{\bibinfo{person}{Saurabh Verma} {and} \bibinfo{person}{Zhi~Li Zhang}.} \bibinfo{year}{2018}\natexlab{}.
\newblock \bibinfo{title}{{Graph capsule convolutional neural networks}}.
\newblock
\newblock
\showISSN{23318422}
\showeprint[arxiv]{1805.08090}


\bibitem[Wang et~al\mbox{.}(2016)]%
        {7462285}
\bibfield{author}{\bibinfo{person}{Xinyu Wang}, \bibinfo{person}{Tsan-Ming Choi}, \bibinfo{person}{Haikuo Liu}, {and} \bibinfo{person}{Xiaohang Yue}.} \bibinfo{year}{2016}\natexlab{}.
\newblock \showarticletitle{Novel Ant Colony Optimization Methods for Simplifying Solution Construction in Vehicle Routing Problems}.
\newblock \bibinfo{journal}{\emph{IEEE Transactions on Intelligent Transportation Systems}} \bibinfo{volume}{17}, \bibinfo{number}{11} (\bibinfo{year}{2016}), \bibinfo{pages}{3132--3141}.
\newblock
\urldef\tempurl%
\url{https://doi.org/10.1109/TITS.2016.2542264}
\showDOI{\tempurl}


\bibitem[Wei et~al\mbox{.}(2021)]%
        {9482700}
\bibfield{author}{\bibinfo{person}{Qinshuang Wei}, \bibinfo{person}{Gustav Nilsson}, {and} \bibinfo{person}{Samuel Coogan}.} \bibinfo{year}{2021}\natexlab{}.
\newblock \showarticletitle{Scheduling of Urban Air Mobility Services with Limited Landing Capacity and Uncertain Travel Times}. In \bibinfo{booktitle}{\emph{2021 American Control Conference (ACC)}}. \bibinfo{pages}{1681--1686}.
\newblock
\urldef\tempurl%
\url{https://doi.org/10.23919/ACC50511.2021.9482700}
\showDOI{\tempurl}


\bibitem[Zhang et~al\mbox{.}(1999)]%
        {zhang1999team}
\bibfield{author}{\bibinfo{person}{Tiehua Zhang}, \bibinfo{person}{WA Gruver}, {and} \bibinfo{person}{Michael~H Smith}.} \bibinfo{year}{1999}\natexlab{}.
\newblock \showarticletitle{Team scheduling by genetic search}. In \bibinfo{booktitle}{\emph{Intelligent Processing and Manufacturing of Materials, 1999. IPMM'99. Proceedings of the Second International Conference on}}, Vol.~\bibinfo{volume}{2}. IEEE, \bibinfo{pages}{839--844}.
\newblock


\bibitem[Zhou et~al\mbox{.}(2021)]%
        {zhou2021informer}
\bibfield{author}{\bibinfo{person}{Haoyi Zhou}, \bibinfo{person}{Shanghang Zhang}, \bibinfo{person}{Jieqi Peng}, \bibinfo{person}{Shuai Zhang}, \bibinfo{person}{Jianxin Li}, \bibinfo{person}{Hui Xiong}, {and} \bibinfo{person}{Wancai Zhang}.} \bibinfo{year}{2021}\natexlab{}.
\newblock \showarticletitle{Informer: Beyond efficient transformer for long sequence time-series forecasting}. In \bibinfo{booktitle}{\emph{Proceedings of the AAAI conference on artificial intelligence}}, Vol.~\bibinfo{volume}{35}. \bibinfo{pages}{11106--11115}.
\newblock


\end{thebibliography}

\end{document}